\documentclass{elsart}
\usepackage{amsmath,amssymb,graphicx}

\begin{document}
\begin{frontmatter}

\title{Towards a model-independent low momentum nucleon-nucleon interaction}

\author{S.K. Bogner$\hspace{1mm}^{a}$\thanksref{SKB}},
\author{T.T.S. Kuo$\hspace{1mm}^{a}$\thanksref{TTSK}},
\author{A. Schwenk$\hspace{1mm}^{a}$\thanksref{AS}},
\author{D.R. Entem$\hspace{1mm}^{b}$ and R. Machleidt$\hspace{1mm}^{b}$}
\address{$^{a}$Department of Physics and Astronomy, State University of
New York,\\
Stony Brook, NY 11794-3800\\
$^{b}$Department of Physics, University of Idaho, P.O. Box 440903,\\
Moscow, ID 83844-0903}

\thanks[SKB]{E-mail: bogner@phys.washington.edu}
\thanks[TTSK]{E-mail: kuo@nuclear.physics.sunysb.edu}
\thanks[AS]{E-mail: aschwenk@mps.ohio-state.edu}

\begin{abstract}

\noindent
We provide evidence for a high precision model-independent 
low momentum nucleon-nucleon interaction. Performing a 
momentum-space renormalization group decimation, we
find that the effective interactions constructed from various high 
precision nucleon-nucleon interaction models, such as the Paris,
Bonn, Nijmegen, Argonne, CD Bonn and Idaho potentials, are identical.
This model-independent low momentum interaction, called $V_{\text{low k}}$,
reproduces the same phase shifts and deuteron pole as the input potential 
models, without ambiguous assumptions on the high momentum components,
which are not constrained by low energy data and lead to model-dependent 
results in many-body applications. $V_{\text{low k}}$ is energy-independent 
and does not necessitate the calculation of the Brueckner $G$ matrix.

\vspace{0.5cm}

\end{abstract}
\end{frontmatter}

\section{Introduction}

In low energy nuclear systems such as finite nuclei 
and nuclear matter, one can base complicated many-body calculations
on a simple picture of point-like nucleons interacting by means of a
two-body potential and three-body forces when needed. Unlike for
electronic systems where the low energy Coulomb force is unambiguously
determined from Quantum Electrodynamics, there is much ambiguity
in nuclear physics owing to the non-perturbative nature of Quantum
Chromodynamics (QCD) at low energy scales. Consequently, there are a
number of high precision, phenomenological meson exchange models of 
the two-nucleon force $V_{\text{NN}}$, such as the
Paris~\cite{Lacombe:1980dr}, Bonn~\cite{Machleidt:1987hj},
Nijmegen~\cite{Stoks:1994}, Argonne~\cite{Wiringa:1995wb}
and CD Bonn~\cite{CDBonn1,CDBonn2} potentials, as well as 
model-independent but less accurate treatments based on chiral 
Effective Field Theory 
(EFT)~\cite{Ordonez:1994tn,KSW,Park:1998kp,Epelbaum:1999dj}, 
for a review see~\cite{EFT:2000}. We also study the 
Idaho potential~\cite{Entem:2001}, which is based on the EFT framework,
but some model dependence is introduced in order to achieve similarly
high precision as compared to the other interaction models.

These nuclear force models incorporate the same one-pion
exchange interaction (OPE) at long distance, but differ
in their treatment of the intermediate and short-range parts (e.g., 
parameterization of the repulsive core compared to $\omega$ 
exchange, different form factors, dispersive or field
theoretical treatment of the $2 \pi$ exchange). The fact that the
different short distance 
constructions reproduce the same phase shifts and deuteron
properties indicates that low energy observables are insensitive
to the details of the short distance dynamics. The EFT approach exploits
this insensitivity by explicitly keeping only the pion and nucleon
degrees of freedom (in accordance with the spontaneous breaking of
chiral symmetry in the QCD vacuum) and encoding the effects of 
the integrated heavy degrees of freedom in the form of couplings which
multiply model-independent delta functions and their 
derivatives, see e.g.~\cite{Lepage:1997cs}. The EFT thus provides a
model-independent description of the two-nucleon system. However, the
high precision, i.e., $\chi^2/\text{datum} \approx 1$, description of
the nucleon-nucleon scattering data provided by conventional models is
at present not achieved by the rigorous EFT potentials.

\begin{figure}[t]
\begin{center}
\includegraphics[scale=0.36,clip=]{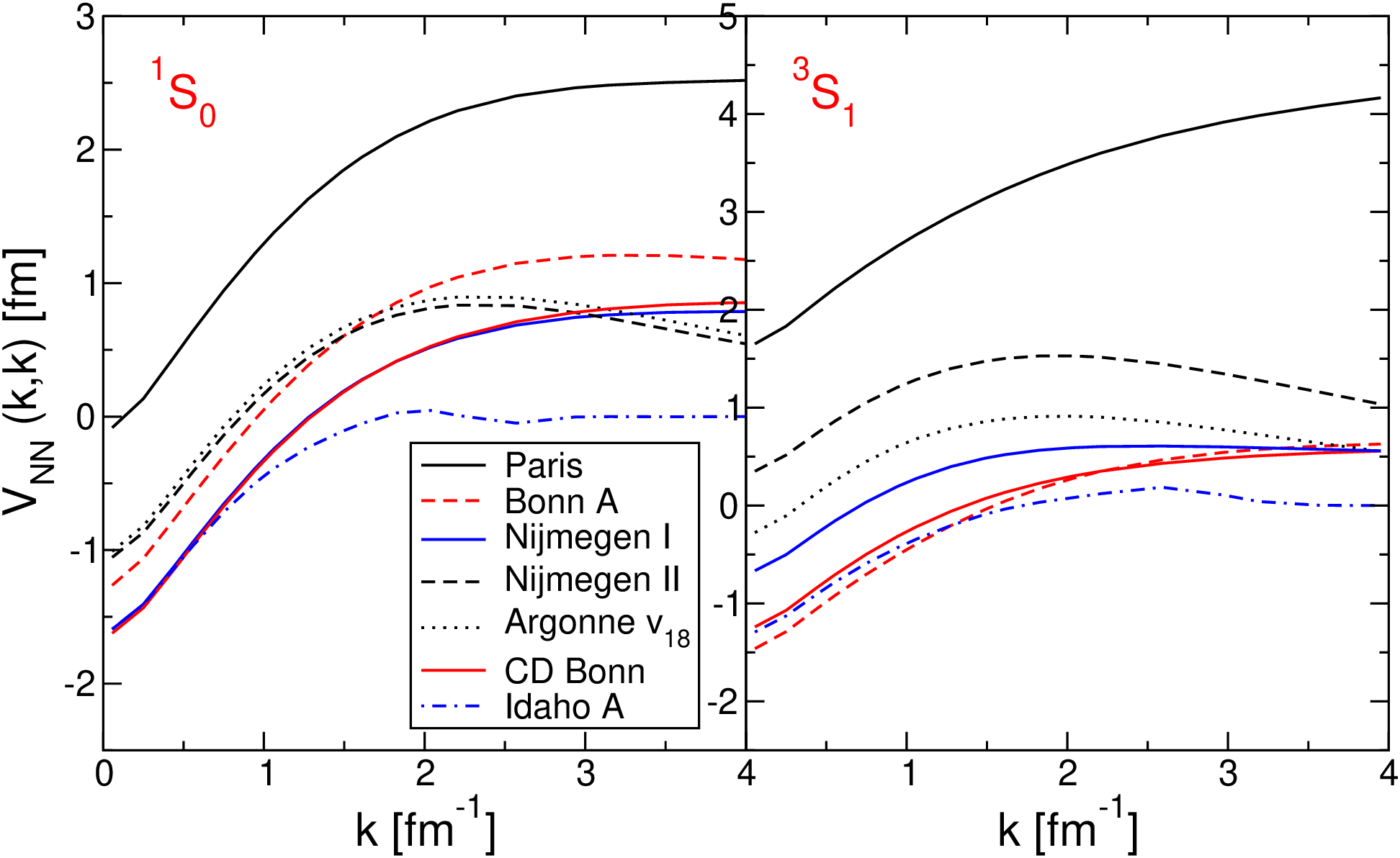}
\end{center}
\caption{Momentum-space matrix elements of $V_{\text{NN}}$ for
different bare potentials in the $^1$S$_0$ and $^3$S$_1$ channels.}
\label{barediag}
\end{figure}
In Fig.~\ref{barediag}, we observe that the realistic models of 
$V_{\text{NN}}$ have quite different momentum-space matrix 
elements despite their common OPE parts and reproduction of 
the same low energy data. It demonstrates that the low energy 
phase shifts and deuteron properties cannot distinguish between the
models used for the short distance parts. However, in many-body
calculations the assumed short-distance structure of a particular 
$V_{\text{NN}}$ enters by means of virtual nucleon states extending to high
momentum, which will lead to model-dependent results, e.g., in
Brueckner Hartree-Fock calculations of the binding energy of nuclear 
matter. It is clearly of great interest to remove such model 
dependence from microscopic nuclear many-body calculations. 

\begin{figure}[t]
\begin{center}
\includegraphics[scale=0.36,clip=]{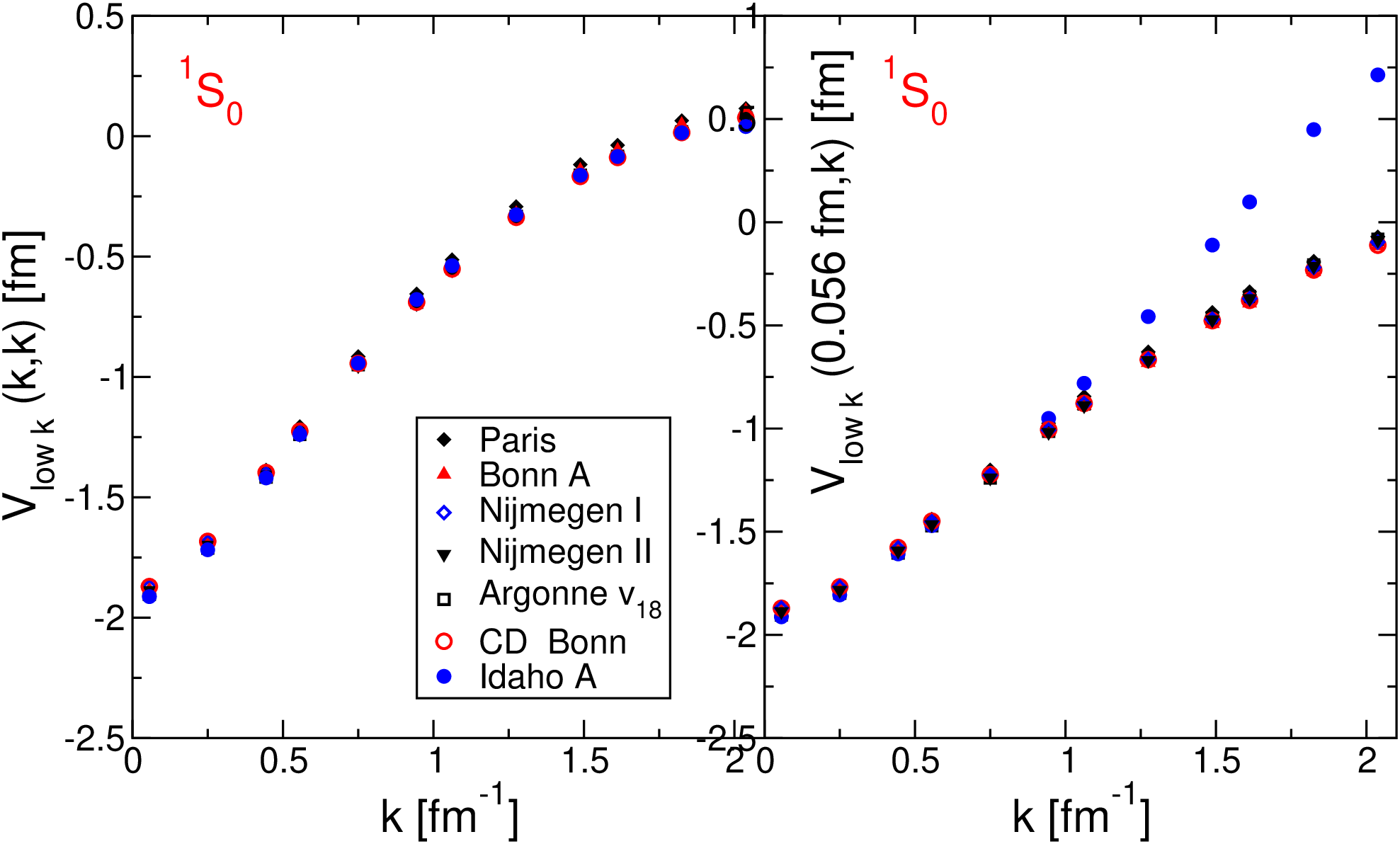}
\end{center}
\caption{Diagonal and off-diagonal momentum-space matrix elements 
of the $V_{\text{low k}}$ obtained from the different bare 
potentials in the $^1$S$_0$ channel for a cutoff $\Lambda = 2.1
\; \text{fm}^{-1}$.}
\label{vlowk}
\end{figure}

Motivated by these observations, we perform a renormalization group
(RG) decimation and integrate out the ambiguous high momentum
components of the realistic interactions. Due to the
separation of scales in the nuclear problem, it is reasonable to
expect that the model dependence of the input potentials will be largely
removed as the high momentum components are excluded 
from the Hilbert space. Starting from 
any of the $V_{\text{NN}}$ in Fig.~\ref{barediag}, we integrate out 
the momenta above a cutoff $\Lambda$ to obtain an effective low 
momentum potential, called $V_{\text{low k}}$. The physical condition is
that the effective theory reproduces the deuteron pole and the 
half-on-shell (HOS) $T$ matrices  of the input $V_{\text{NN}}$ 
model (i.e., the observable scattering phase shifts and the 
low momentum components of the two-body wave functions, which 
probe the OPE part of the interaction), but with all loop momenta cut off
at $\Lambda$. We will find the striking result of Fig.~\ref{vlowk} 
that the $V_{\text{low k}}$ becomes independent of the particular 
input model for $\Lambda \lesssim 2.1 \;\text{fm}^{-1}$, which
corresponds to laboratory energies $E_{\text{lab}} \lesssim 350 \; 
\text{MeV}$. The latter is the energy scale over which the high
precision potentials are constrained by experimental data.

Our strategy thus is a compromise between conventional models and EFT
treatments. The resulting $V_{\text{low k}}$ reproduces the
nucleon-nucleon scattering phase shifts with similar accuracy as the
high precision potentials, but without making further assumptions on the 
detailed high momentum structure, which cannot be resolved by fitting to
the low energy data only.

\section{Renormalization group decimation}

The first step in the RG decimation is to define an 
energy-dependent effective potential (called the $\widehat{Q}$ box or
the Bloch-Horowitz potential in effective interaction theory), 
which is irreducible with respect to cutting 
intermediate low momentum propagators. The
$\widehat{Q}$ box resums the effects of the high momentum modes,
\begin{equation}
\widehat{Q}(k',k;\omega) = V_{\text{NN}}(k',k) + \frac{2}{\pi} \;
\mathcal{P} \int_{\Lambda}^{\infty} \frac{V_{\text{NN}}(k',p) \;
\widehat{Q} (p,k;\omega)}{\omega-p^{2}} \; p^{2} dp , \label{Qbox}
\end{equation}
where $k'$, $k$, and $p$ denote the relative momentum of the
outgoing, incoming, and intermediate nucleons. The principal value
HOS $T$ matrix for a given partial wave is obtained by solving the
Lippmann-Schwinger equation
\begin{equation}
T(k',k;k^{2}) = V_{\text{NN}}(k',k) +  \frac{2}{\pi}\; \mathcal{P}
\int_{0}^{\infty} \frac{V_{\text{NN}}(k',p) \;
T(p,k;k^{2})}{k^{2}-p^{2}} \; p^{2} dp. \label{Tbare}
\end{equation}
In terms of the effective cutoff-dependent $\widehat{Q}$ potential, 
the scattering equation can be expressed as
\begin{equation}
T(k',k;k^{2}) = \widehat{Q}(k',k;k^{2}) + \frac{2}{\pi} \;
\mathcal{P} \int_{0}^{\Lambda} \frac{\widehat{Q}(k',p;k^{2}) \;
T(p,k;k^{2})}{k^{2}-p^{2}} \; p^{2} dp . \label{TQeq}
\end{equation}
The low momentum effective theory defined by Eq.~(\ref{Qbox}) 
and Eq.~(\ref{TQeq}) preserves the low energy scattering
amplitudes and bound states independently of the chosen model space,
i.e., the value of the cutoff $\Lambda$. However, the energy 
dependence of the effective $\widehat{Q}$ potential is 
inconvenient for practical calculations. In order to eliminate
the energy dependence, we introduce the so-called Kuo-Lee-Ratcliff
folded diagrams, which provide a way of reorganizing the Lippmann-Schwinger
equation, Eq.~(\ref{TQeq}), such that the energy dependence of the
$\widehat{Q}(k',k;\omega)$ box is converted to a purely momentum
dependent interaction, $V_{\text{low k}}(k',k)$~\cite{KLR,Kuo:1990} .
The folded diagrams are correction terms one must add to
Eq.~(\ref{TQeq}), if one were to set all $\widehat{Q}$ box
energies right side on-shell. This explicitly leads to (for
details see~\cite{Kuo:1990,TmatrixRG})
\begin{multline}
V_{\text{low k}}(k',k) = \widehat{Q}(k',k;k^{2}) + \; \frac{2}{\pi} \;
\mathcal{P} \int_{0}^{\Lambda} p^2 dp \; \widehat{Q}(p,k;k^{2}) \\
\times \frac{\widehat{Q}(k',p;k^{2}) -
\widehat{Q}(k',p;p^{2})}{k^{2}-p^{2}} \: + \: \mathcal{O}(\widehat{Q}^3) .
\label{VfromQ}
\end{multline}
The folded diagram resummation indicated in Eq.~(\ref{VfromQ}) 
can be carried out to all orders using the similarity transformation 
method of Lee and Suzuki~\cite{LS1,LS2}. By construction, the 
resulting energy-independent $V_{\text{low k}}$ preserves 
the low momentum HOS $T$ matrix of the input $V_{\text{NN}}$
model~\cite{Vlowk},
\begin{equation}
T(k',k;k^{2}) = V_{\text{low k}}(k',k) +  \frac{2}{\pi} \;
\mathcal{P} \int_{0}^{\Lambda} \frac{V_{\text{low k}}(k',p) \;
T(p,k;k^{2})}{k^{2}-p^{2}} \; p^{2} dp ,
\end{equation}
where all momenta are constrained to lie below the cutoff $\Lambda$. 
As our RG decimation preserves the half-on-shell $T$ matrix,
we have $d T(k',k;k^{2}) / d \Lambda = 0$, which implies a RG
equation for $V_{\text{low k}}$~\cite{TmatrixRG}
\begin{equation}
\frac{d}{d \Lambda} V_{\text{low k}}(k',k) =
\frac{2}{\pi} \frac{V_{\text{low k}}(k',\Lambda) \;
T(\Lambda,k;\Lambda ^{2})}{1-(k / \Lambda)^{2}} .
\label{flow}
\end{equation}
Similarly, a scaling equation is obtained for the $\widehat{Q}$ box 
by integrating out an infinitesimal momentum shell, and one finds
\begin{equation}
\frac{d}{d \Lambda} \widehat{Q}(k',k;p^{2}) =
\frac{2}{\pi}\frac{\widehat{Q}(k',\Lambda;p^{2}) \;
\widehat{Q}(\Lambda,k;p^{2})}{1-(p / \Lambda)^2} .
\label{Qflow}
\end{equation}
This equation was obtained previously by Birse {\it et al.} by
requiring the invariance of the full-off-shell $T$
matrix, $d T(k',k;p^2) / d \Lambda = 0$~\cite{Birse:1998dk}.

The RG equation, Eq.~(\ref{flow}), lies at the heart 
of the approach presented here: given
a microscopic input model with a large cutoff, one can use the RG 
equation to evolve the bare interaction to a physically equivalent, but
simpler effective theory valid for energies below the cutoff. The RG
evolution separates the details of the assumed short distance
dynamics, while incorporating their detail-independent effects on
low energy phenomena through the running of the effective
interaction~\cite{TmatrixRG,Bogner:2000js}.

The principal difference between the presented RG approach and the 
standard EFT one is that we do not expand the interaction in powers 
of local operators. This implies that, first, one does not make
assumptions on locality, and that second, we do not truncate after 
a certain power included in the low momentum interaction.
In the EFT approach, power-counting is used to calculate observables 
to a given order with controllable errors. In contrast, we start
from a Hamiltonian in a large space, which reproduces the low-energy
observables with high accuracy and then decimate to a smaller, low 
momentum space so that the observables are reproduced. Since the exact 
RG equation is solved without truncation, the invariance of the $T$ 
matrices is guaranteed.

Nevertheless, there are close similarities between the presented RG 
approach and the standard EFT one. For example, at cutoffs above
the pion mass, $V_{\text{low k}}$ keeps the pion exchange explicit, while
the unresolved short distance physics could be encoded in a series of 
contact terms. Moreover, we find below that $V_{\text{low k}}$ seems only 
to depend on the fact that all potential models have the same pion tail, and 
fit the same phase shifts. This is similar to EFT treatments, where the
interaction is constrained by pion exchange, phase shifts and the choice 
of regulator. 

\section{Results}

\begin{figure}[t]
\begin{center}
\includegraphics[scale=0.54,clip=]{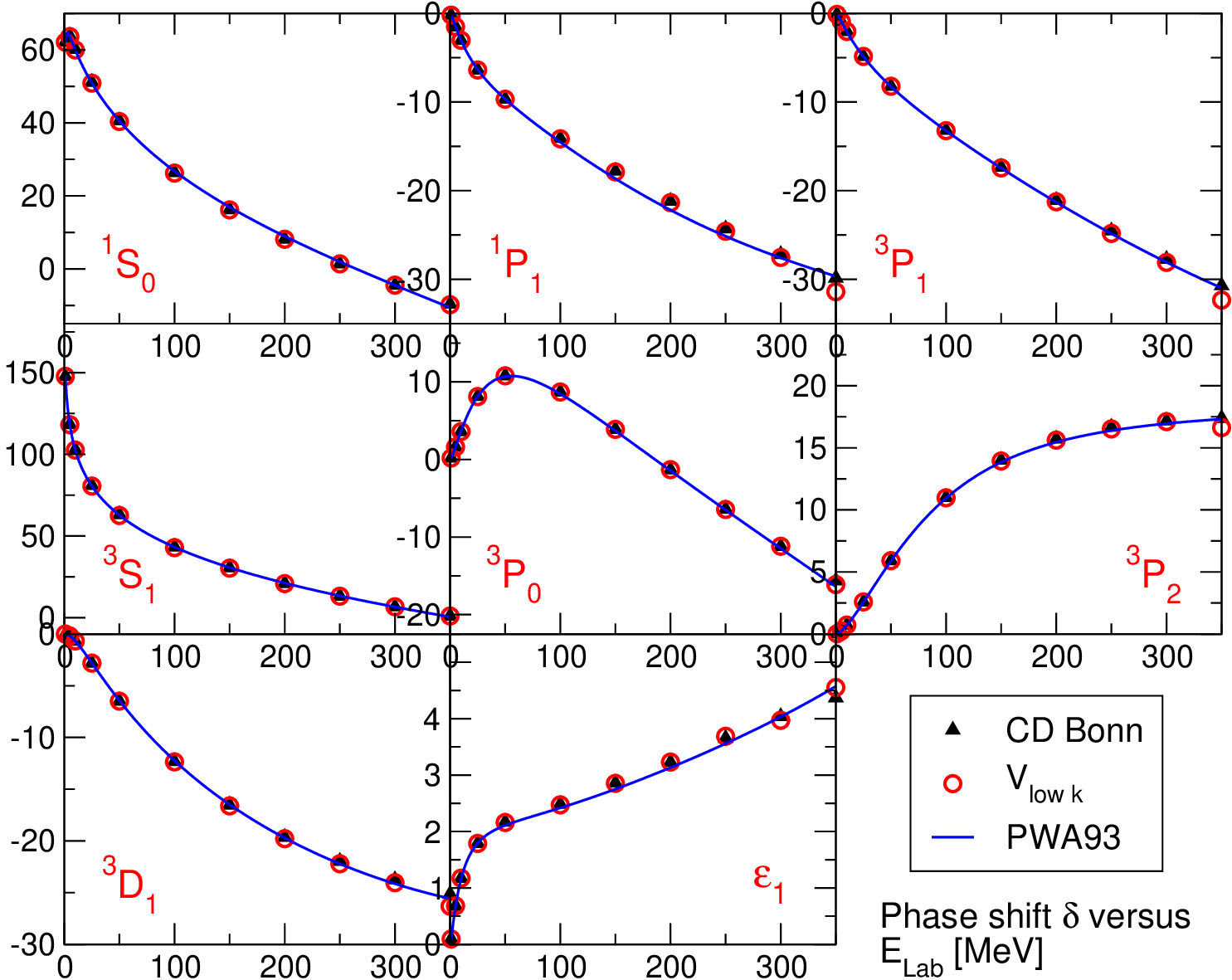}
\end{center}
\caption{S-wave (singlet and triplet with mixing parameter) and 
P-wave phase shifts of $V_{\text{low k}}$ for a cutoff $\Lambda = 2.1
\; \text{fm}^{-1}$ compared to the input $V_{\text{NN}}$ model 
used, here the CD Bonn potential. We also show the results of the 
multi-energy phase shift analysis (PWA93) of the Nijmegen group~\cite{PSA}.
For $V_{\text{low k}}$ the agreement of the calculated phase shifts 
with the PWA93 is determined by the quality of the fit for the bare
interaction.}
\label{phaseshifts}
\end{figure}

Referring to Fig.~\ref{vlowk}, we find the central result of 
this Letter. The diagonal matrix elements of $V_{\text{low k}}$ 
obtained from the different $V_{\text{NN}}$ of Fig.~\ref{barediag} 
collapse onto the same curve for $\Lambda \lesssim 2.1 
\;\text{fm}^{-1}$. Similar results are found in all partial 
waves and will be reported elsewhere in detail~\cite{bigVlowk}. 
We emphasize that $V_{\text{low k}}$ reproduces the experimental 
phase shift data and the deuteron pole with similar accuracy as 
compared with the high precision potential models, as we show in
Fig.~\ref{phaseshifts}. The reproduction of the phase 
shifts with the renormalized $V_{\text{low k}}$ does however not
require ambiguous high momentum components, which are not 
constrained by the low energy scattering data. We note that the 
change in sign of the $^1$S$_0$ and the $^3$P$_0$ phase shifts 
indicates that the effects of the repulsive core are properly 
encoded in $V_{\text{low k}}$.

Intriguingly, the $V_{\text{low k}}$ mainly differs from the bare
potential by a constant shift. This was previously 
observed by Epelbaum {\it et al.} for a toy two-Yukawa bare
potential~\cite{Epelbaum:1998na}. The constant shift in momentum
space corresponds to a smeared delta function in coordinate space and
accounts for the renormalization of the repulsive core from the bare
interactions, see also~\cite{Neff}.

In Fig.~\ref{vlowk}, we find a similar behaviour for the
off-diagonal matrix elements, although the $V_{\text{low k}}$ derived
from the Idaho potential begins to differ from the others
at approximately $2 \, m_\pi = 1.4 \; \text{fm}^{-1}$. This
discrepancy in the off-diagonal matrix elements arises from the fact that the
Idaho potential used here treats the 2$\pi$ exchange differently 
than the meson models do. We can integrate out further and lower 
the cutoff to $\Lambda \lesssim 1.4 \; \text{fm}^{-1}$, then the
off-diagonal elements collapse as well.

\begin{figure}[t]
\begin{center}
\includegraphics[scale=0.44,clip=]{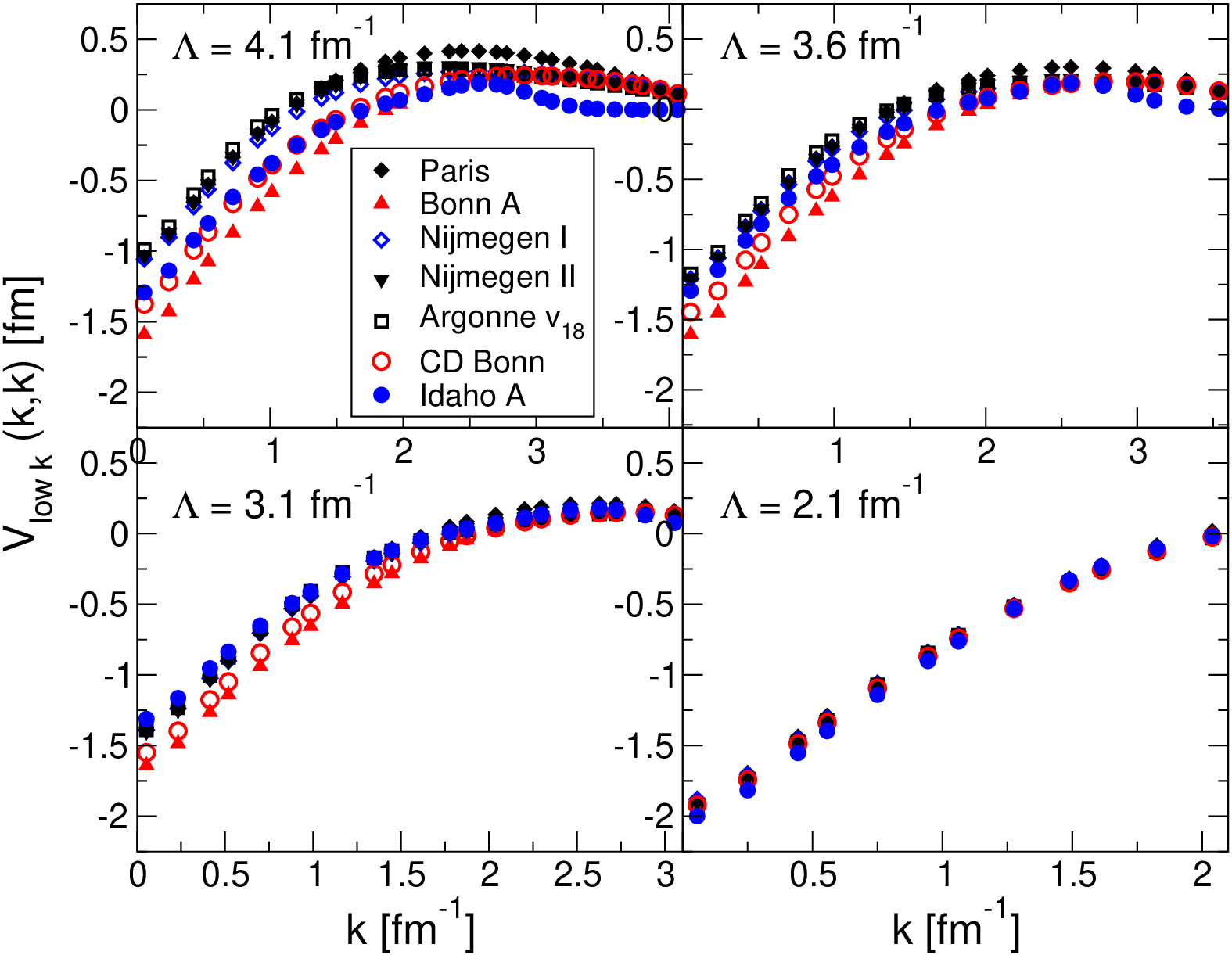}
\end{center}
\caption{Evolution of the diagonal momentum-space matrix elements for 
the $V_{\text{low k}}$ derived from the different bare potentials
from larger values of the cutoff to $\Lambda = 2.1 \; \text{fm}^{-1}$ 
in the $^3$S$_1$ partial wave.}
\label{3s1collaps}
\end{figure}
These results can be understood from $T$ matrix preservation.
The HOS $T$ matrix determines the phase shifts as well as the 
low momentum components of the low energy scattering and bound state wave
functions. Using the spectral representation of the $T$ matrix,
$V_{\text{low k}}$ can be expressed as
\begin{equation}
V_{\text{low k}}(k^\prime,k) =  T(k^\prime,k;k^2) \\
+ \; \frac{2}{\pi} \; \mathcal{P} \int_{0}^{\Lambda}  T(k^\prime,p;p^2) \; \frac{1}{p^2 - k^2} \; T(p,k;p^2) .
\label{VfromT}
\end{equation}
The $V_{\text{NN}}$ models give the same on-shell $T$ matrices
over the phase shift equivalent kinematic range of $k \lesssim 
2.1 \;\text{fm}^{-1}$, but their off-shell behavior is a priori
unconstrained. In practice, however, one observes that the realistic
potential models result in similar HOS $T$ matrices at low energies 
and momenta. This is understood from the fact that these models share 
the same long-range OPE interaction and differ most noticeably on
short-distance scales set by the repulsive core, $r \sim 0.5
\;\text{fm}$ and smaller. Consequently, one expects that the off-shell 
differences are suppressed at energies and momenta below a
corresponding scale of $\Lambda \sim 1/r \sim 2.0 \;\text{fm}^{-1}$. 
It is clear from Eq.~(\ref{VfromT}) that the approximate HOS $T$ 
matrix equivalence is a sufficient condition for the $V_{\text{low k}}$ 
to be independent of the various potential models. 
Moreover, the spectral representation clarifies that the off-diagonal 
matrix elements of $V_{\text{low k}}$ are more sensitive to a particular
off-shell behavior as observed in Fig.~\ref{vlowk}, and thus deviate at 
a lower cutoff than the diagonal matrix elements. The collapse of the
diagonal momentum-space matrix elements at the scale set by the
constraining scattering data, $\Lambda \approx 2.1 \;\text{fm}^{-1}$,
is nicely illustrated in Fig.~\ref{3s1collaps}, which shows the RG 
evolution as the cutoff is successively lowered in the $^3$S$_1$
partial wave.

\begin{figure}[t]
\begin{center}
\includegraphics[scale=0.36,clip=]{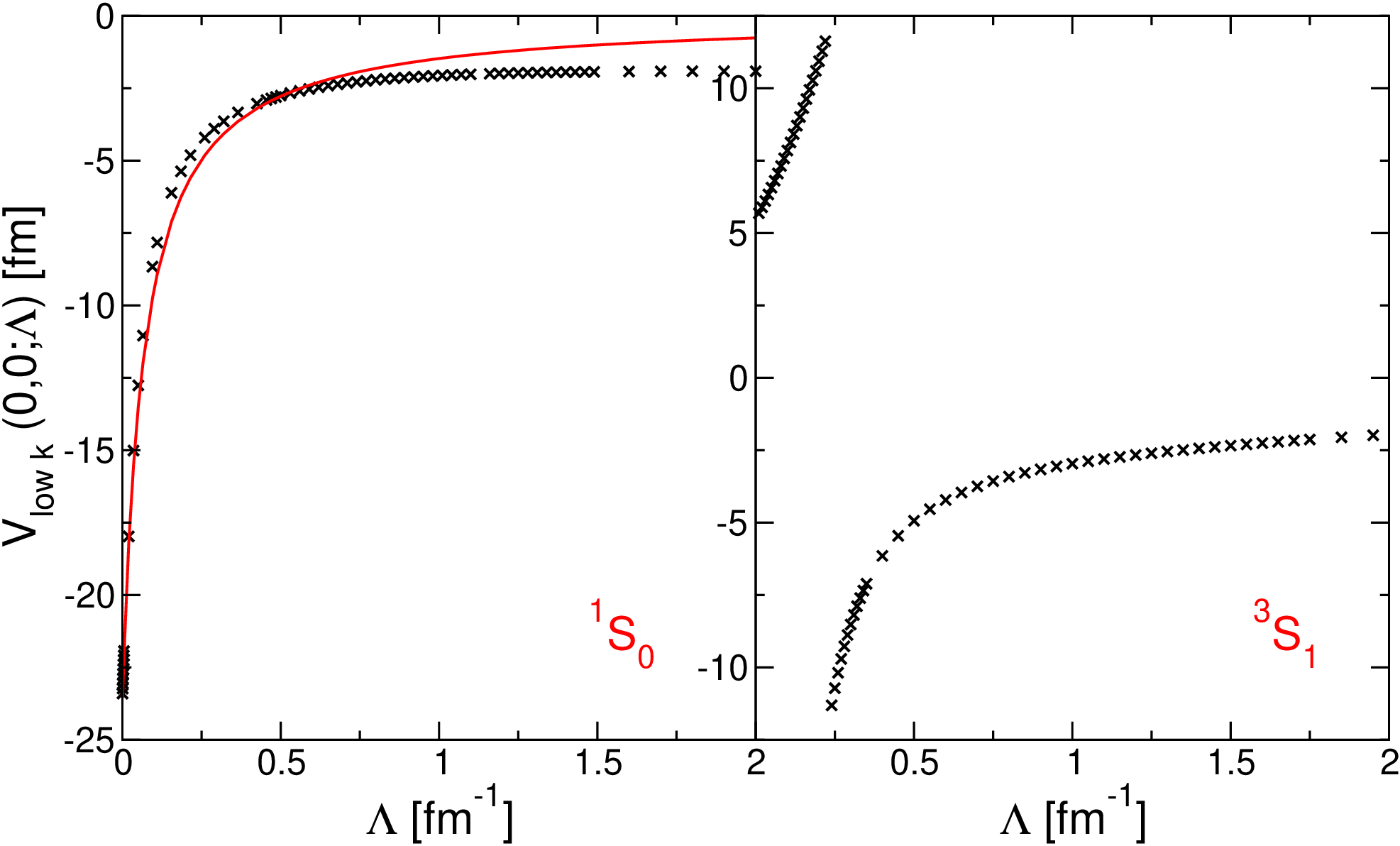}
\end{center}
\caption{RG flow of $V_{\text{low k}}(0,0;\Lambda)$ versus cutoff
$\Lambda$ in the $^1$S$_0$ and $^3$S$_1$ partial waves. The solid 
line represents the solution of the RG equation for small $\Lambda$
as discussed in the text.}
\label{swavelambda}
\end{figure}

Next, we analyze the scaling properties of $V_{\text{low k}}$.
For this purpose, we show the matrix element $V_{\text{low k}}(0,0)$ 
versus cutoff in Fig.~\ref{swavelambda} for the $^1$S$_0$ and $^3$S$_1$ 
partial waves. The main results are the following.
$T$ matrix preservation guarantees that $V_{\text{low k}}(0,0)$
flows toward the scattering length as the cutoff is taken to zero, these are
$a_{^1\text{S}_0} = - 23.73 \; \text{fm}$ and $a_{^3\text{S}_1} = 5.42 \;
\text{fm}$ and are well reproduced with $V_{\text{low k}}$.

For cutoffs $\Lambda > m_{\pi}$, $V_{\text{low k}}$ is nearly
independent of the cutoff in the $^1$S$_0$ channel and weakly linearly
dependent in the $^3$S$_1$ channel. According to EFT principles,
the couplings are nearly independent of the cutoff as long as
$\Lambda$ is large enough to explicitly include the relevant
degrees of freedom needed to describe the
scale one is probing. Thus, the running of $V_{\text{low k}}$
initiated at $\Lambda \sim m_{\pi}$ is a result of integrating out
the pion. The rapid changes at very small $\Lambda$ are the result
of the large scattering length. To illustrate the effects of the
large scattering length, we solve the RG equation for
$V_{\text{low k}}(0,0)$ for small cutoffs, with the $^1$S$_0$
scattering length as the $\Lambda=0$ boundary condition. One finds
\begin{equation}
V_{\text{low k}}(0,0) \approx \frac{1}{1 / a_{^1\text{S}_0}  -
2 \Lambda / \pi} ,
\label{smalllambda}
\end{equation}
which is shown as the solid curve in Fig.~\ref{swavelambda}. Clearly, 
the agreement for small $\Lambda < m_\pi/2$ is
convincing. Eq.~(\ref{smalllambda}) is identical to the
Kaplan-Savage-Wise solution for the renormalization of the
momentum-independent contact term in the pionless EFT, upon
identifying $\Lambda$ with the regulator mass $\mu$ used 
in dimensional regularization~\cite{KSW}. The weak dependence on 
the cutoff in the $^3$S$_1$ partial wave results from the dominantly 
second order tensor contributions, which are peaked at relative 
momenta $k \approx 2 \; \text{fm}^{-1}$~\cite{UTNMF}.

Finally, we note that $V_{\text{low k}}$ preserves 
all scattering and bound states with
energy $|E| < \Lambda^2$. Therefore, the jump in the $^3$S$_1$
channel is the bound state contribution to the $T$ matrix
and occurs at $\Lambda = k_{\text{D}} = \sqrt{m \,
E_{\text{D}}/\hbar^2} = 0.23 \; \text{fm}^{-1}$. From effective
range theory and the Low equation~\cite{NNint}, we have for the bound
state contribution and thus the discontinuity
\begin{align}
\Delta V_{\text{low k}}(0,0;k_{\text{D}})_{^3\text{S}_1} =\frac{4}{\pi \,
k_{\text{D}} \, ( 1 - k_{\text{D}} \, r_0 )} = 22.8 \;
\text{fm} ,
\end{align}
where $r_0 = 1.75 \; \text{fm}$ denotes the effective range in
$^3$S$_1$. This is in very good agreement with our results.

\section{Summary}

In this work, we have shown that the separation of scales in the
nuclear force can be successfully applied to derive a model-independent 
low momentum nucleon-nucleon interaction. $V_{\text{low k}}$ is 
obtained by integrating out the high momentum 
components of various potential models, leading to a
physically equivalent effective theory in the low momentum Hilbert 
space. The decimation filters the details of the assumed
short-distance dynamics of the bare interactions, provided one 
requires that the low energy observables are preserved under the RG. 
We have argued and provided numerical evidence that the model-independence 
of $V_{\text{low k}}$ is a consequence of the common long-range OPE interaction
and the reproduction of the same elastic scattering phase shifts.
The momentum scale $\Lambda \sim 2.1 \;\text{fm}^{-1}$, corresponding 
to laboratory energies $E_{\text{lab}} \sim 350 \; \text{MeV}$, is
precisely the scale at which the low momentum interaction becomes
independent of the input models. Our work demonstrates that the 
differences in the high momentum components of the nucleon force are
not constrained by fits to the low energy phase shifts and the 
deuteron properties.

$V_{\text{low k}}$ does not have high momentum modes, which are 
related to the strong, short range-repulsion in conventional 
models. Therefore, the low momentum interaction is considerably 
softer than the bare interactions (both in a plane-wave and 
an oscillator basis) and does not have a repulsive core.
As a consequence, when $V_{\text{low k}}$ is used as the 
microscopic input in the many-body problem, the high momentum 
effects in the particle-particle channel do not have to be 
addressed by performing a Brueckner ladder resummation or 
short-range correlation methods. In fact, with a cutoff on
relative momenta, the phase space in the particle-particle
channel is comparable to the particle-hole channels, and it
would seem strange to resum the particle-particle 
channel, while the particle-hole channels are treated 
perturbatively in a hole-line expansion.

The use of $V_{\text{low k}}$ in microscopic nuclear many-body
calculations leads to model-independent results. $V_{\text{low k}}$ 
has been successfully used as shell model effective interaction 
in model space calculations for two valence particle nuclei 
such as $^{18}$O and $^{134}$Te~\cite{Bogner:2001}. 
The starting point of these calculations has traditionally 
been the Brueckner $G$ matrix, which depends on the bare 
$V_{\text{NN}}$ used as well as the particular nuclei 
via the Pauli blocking operator. By means of $V_{\text{low k}}$, 
the same low momentum interaction is used in different mass
regions. $V_{\text{low k}}$ has further been incorporated 
into Fermi liquid theory. This connects the low momentum 
interaction in free space and the quasiparticle interaction in 
normal Fermi systems. Two constraints have been derived
which relate the Fermi liquid parameters of nuclear matter
to the S-wave low momentum interaction at zero relative 
momentum~\cite{Schwenk:2002}. Finally, adopting the RG 
approach to Fermi systems proposed by Shankar, $V_{\text{low k}}$ 
is taken as the input to microscopic calculations of the 
quasiparticle interactions and the pairing gaps in neutron 
matter~\cite{Schwenk:2003}. From the success of these applications, 
we believe that the model-independent $V_{\text{low k}}$ is a very 
promising starting point for microscopic nuclear many-body calculations.

Few-body calculations with $V_{\text{low k}}$ are in progress. 
Preliminary results for the ground state energies of triton and 
$^3$He show that the three-body force for $V_{\text{low k}}$ is 
weaker than e.g., the conventional three-body force constructed 
for the Argonne potential.

\begin{ack}
We thank Gerry Brown for his encouragement and many stimulating discussions.
This work was supported by the US-DOE grant DE-FG02-88ER40388, the
US-NSF grant PHY-0099444 and by the Ramon Areces Foundation of Spain.
\end{ack}

\end{document}